\documentclass[twocolumn,superscriptaddress,amsmath,ammsymb]{revtex4-2}
\usepackage{epsfig}
\usepackage{amsmath}
\usepackage{amssymb}
\usepackage{graphicx}
\usepackage{dcolumn}
\usepackage{bm}
\usepackage[usenames]{color}
\usepackage[breaklinks]{hyperref}
\hypersetup{colorlinks=true,allcolors=blue}

\begin{document}
	
\title{High Chern number phase in topological insulator multilayer structures: a Dirac cone model study}

\author{Yi-Xiang Wang}
\email{wangyixiang@jiangnan.edu.cn}
\affiliation{School of Science, Jiangnan University, Wuxi 214122, China}
\affiliation{School of Physics and Electronics, Hunan University, Changsha 410082, China}

\author{Fuxiang Li}
\email{fuxiangli@hnu.edu.cn}
\affiliation{School of Physics and Electronics, Hunan University, Changsha 410082, China}

\date{\today}	
	
\begin{abstract} 
We use the Dirac cone model to explore the high Chern number $(C)$ phases that are realized in the magnetic-doped topological insulator (TI) multilayer structures by Zhao \textit{et al.} [Nature {\bf588}, 419 (2020)].  The Chern number is calculated by capturing the evolution of the phase boundaries with the parameters and then the Chern number phase diagrams of the TI multilayer structures are obtained.  The high-$C$ behavior is attributed to the band inversion of the renormalized Dirac cones, along with which the spin polarization at the $\Gamma$ point will get increased.  Moreover, another two TI multilayer structures as well as the TI superlattice structures are also studied.  


\end{abstract}  

\maketitle

\section{Introduction} 

The quantum anomalous Hall (QAH) insulator, or called the Chern insulator, has aroused the ongoing interests in condensed matter physics, where the dissipationless chiral currents can flow in a zero magnetic field~\cite{D.J.Thouless, F.D.M.Haldane, X.L.Qi}.  It was first observed in magnetic topological insulator (TI) thin films~\cite{C.Z.Chang2013} that are formed by an intricate interplay between the ferromagnetism due to the magnetic-doping and the intrinsic spin-orbit coupling~\cite{R.Yu}.  The magnetic-doping breaks the time-reversal symmetry (TRS) in the system and can open a small Dirac mass gap in the topological surface states~\cite{Y.L.Chen}.  In these systems, the unavoidable inhomogeneity of the magnetic-doping would lead to the complex magnetic orders~\cite{J.Zhang, X.Kou2015}.  As a consequence, the QAH effect can only be observed at extremely low temperature, which is about several tens of milikelvin, one to two orders of magnitude below the Curie temperature of the magnetic dopants~\cite{C.Z.Chang2013, X.Kou2015, J.G.Checkelsky, Y.Feng, X.Kou2014, C.Z.Chang2015}.  When the crystalline structure and magnetic order are self-organized into a well-ordered topological superlattice, the QAH insulator can be observed at much higher temperature~\cite{Rienks, H.Deng}.  

Besides the $C=1$ Chern insulator phase, a recent experimental work by Zhao \textit{et al.} demonstrated the existence of the high Chern number $C>1$ phase in the TI multilayer structures~\cite{Y.F.Zhao}.  The authors explained the high-$C$ behavior through the interface Dirac state mechanism, where a nontrivial interface state appears at the interface between the magnetic-doped and undoped TI layers and when such a state is occupied, it will contribute $\frac{e^2}{2h}$ to the total Hall conductance $\sigma_{xy}$.  In our previous work~\cite{Y.X.Wang2021}, we instead attributed the inverted bands in the high-$C$ phase to the two-dimensional (2D) subbands, which may not appear at the interface, but are splitted by the thickness confinement of the total stacking layers.  In both theoretical analysis~\cite{Y.F.Zhao, Y.X.Wang2021}, the bulk $\boldsymbol k\cdot\boldsymbol p$ model of the three-dimensional (3D) TIs~\cite{H.Zhang, C.X.Liu} were used to describe the TI multilayer structures and the plane-wave expansion method was adopted to satisfy the open boundary condition at both ends.  The plane-wave expansion method can correctly describe the top and bottom surface states, so it works well for the TI thin film~\cite{C.X.Liu,J.Wang,B.Zhou}.  For the TI multilayers, they were fabricated by the MBE method in the experiment and thus the linear Dirac cones should be present on each surface of the TI layers.  However, this fact may not be effectively captured by the plane-wave expansion method, as the electronic states were found to even penetrate into the deep TI layer interior~\cite{Y.F.Zhao, Y.X.Wang2021}.

On the other hand, as the bulk states of the TIs are gapped, a highly simplified Dirac cone model was initially proposed to explain the QAH behavior in the magnetic TI thin film~\cite{R.Yu}. The model was later developed to describe the magnetic-doped TI and ordinary insulator multilayers that can host the 3D Weyl semimetal (WSM) phase~\cite{A.A.Burkov, A.A.Zyuzin}.  In the Dirac cone model, only the Dirac cone degrees of freedom on each TI surface were retained, with the intralayer and interlayer Dirac cone hoppings being included; while the bulk electronic dynamics were completely ignored.  This was demonstrated to be valid by comparing the model parameters with the DFT band  calculations~\cite{C.Lei}.  Furthermore, the Dirac cone model was used to explore the dependence of the QAH behavior on the film thickness, the magnetic configuration as well as the stacking sequences of the TI layers~\cite{C.Lei}.  
 
Motivated by these progresses, here we try to investigate the high-$C$ behavior in the magnetic-doped TI multilayer structures from the point of view of the Dirac cone model.  We focus on the Chern number phase diagram, where the Chern number is calculated by capturing the evolution of the phase boundaries with the parameters~\cite{Y.X.Wang2021, J.Wang}.  For comparison, we also consider another two TI multilayer structures as well as the 3D TI superlattice structures.  Our main findings are as follows: (i) Within the Dirac cone model, the high-$C$ behavior is attributed to the band inversion of the renormalized Dirac cones that is driven by the exchange splitting, along with which the spin polarization at the $\Gamma$ point gets increased.  In the highest-$C$ phase, the occupied states are fully downspin polarized at the $\Gamma$ point.  (ii) To explain the observation that the Chern number decreases with the decreasing middle magnetic-doped layer thickness~\cite{Y.F.Zhao}, we assume that the related intralayer Dirac cone hopping changes from negative to positive.  Based on this assumption, it is interesting to find that at certain magnetic-doping, the Chern number can also get increased when the layer thickness decreases.  (iii) For another two TI multilayer structures, the band inversion cannot be achieved for all negative-mass Dirac cones even when the magnetic exchange splittings are very strong.  (iv) For the 3D TI superlattice structures, besides the high-$C$ phases, the WSM phases can appear and span broad parameter regions.  Our work can provide more insights on the high-$C$ behavior realized in the TI multilayer structures, which may pave the way for the future topological electronic devices.

\section{Model and method}

\begin{figure}
	\includegraphics[width=6.2cm]{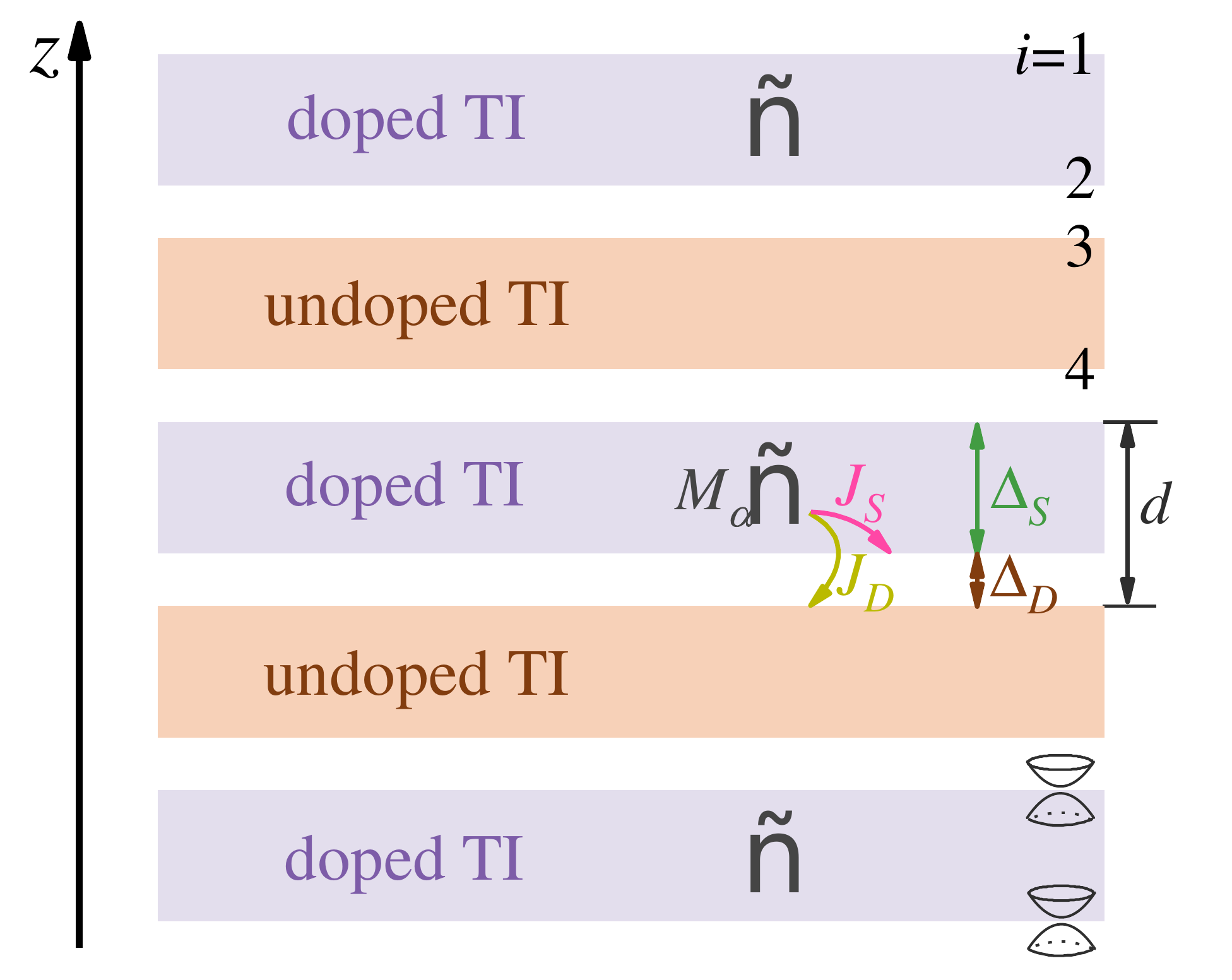}
	\caption{(Color online) Schematics of the TI multilayer structure, including the alternating magnetic-doped ($N_1=3$) and undoped ($N_2=2$) TI layers.  The magnetization in the magnetic-doped TI layer is along the $z$ direction and the interlayer distance is denoted by $d$.  The Dirac cone hoppings $\Delta_S$ and $\Delta_D$, and the exchange splittings $J_S$ and $J_D$ are indicated by the arrows.  Without hoppings and exchange splittings, the Dirac cones are localized on the top and the bottom surface of each TI layer.}
	\label{Fig1}	
\end{figure}

The experimental configuration of the TI multilayers~\cite{Y.F.Zhao} is plotted in Fig.~\ref{Fig1}. It stacks along the $z$ direction and includes the alternating magnetic-doped and undoped TI layers, with the layer number in the plot being $N_1=3$ and $N_2=2$, respectively, so the undoped TI layer is sandwiched between the doped TI layers. 

We use the Dirac cone model to describe the dynamics of the TI multilayer structures.  In the model, the linear Dirac cones are assumed to be present on both the top and bottom surfaces of the magnetic-doped TI layer as well as the undoped TI layer.  The Hamiltonian is written as $(\hbar=1)$~\cite{A.A.Burkov, A.A.Zyuzin, C.Lei}
\begin{align}
H(\boldsymbol k)&=\sum_i\Big[ 
(-1)^i v (-k_x\sigma_y+k_y\sigma_x) 
+m_i \sigma_z \Big]
c_{\boldsymbol k i}^\dagger c_{\boldsymbol k i}
\nonumber\\
&+\sum_{i\in\text{odd}, j}(\Delta_S \delta_{i,j-1}
+\Delta_D \delta_{i+1,j} + \Delta_D \delta_{i-1,j})
c_{\boldsymbol k i}^\dagger c_{\boldsymbol k j}.  
\label{Hk}
\end{align}  
Here $c_{\boldsymbol k i}$ $(c_{\boldsymbol k i}^\dagger)$ annihilates (creates) a surface Dirac cone with the wavevector $\boldsymbol k=(k_x,k_y)$ and the surface index $i$.  For $i$, we use the odd/even number to represent the top/bottom surface of each layer, respectively.  $v$ is the velocity of the Dirac cone, $\sigma$ denote the Pauli matrice that act on the spin space, $\Delta_S$ is the intralayer Dirac cone hopping, and $\Delta_D$ the interlayer Dirac cone hopping across the van der Waals gap.  When the mass term $m_i$ is absent, the system owns the TRS as ${\cal T}H(-\boldsymbol k){\cal T}^{-1}=H(\boldsymbol k)$, with the time-reversal operator ${\cal T}=i\sigma_y K$ and $K$ being the complex conjugate operator. The mass term $m_i$ of each Dirac cone that is induced by the exchange interactions with the nearest local magnetic moments can break the TRS of the system,
\begin{align}
m_i=\sum_\alpha J_{i\alpha}M_\alpha, 
\end{align} 
where $\alpha$ is the layer index.  We assume that the magnetization is along the $z$ direction, so that $M_\alpha=1$ in the magnetic-doped layer, and $M_\alpha=0$ in the undoped layer.  As shown in Fig.~\ref{Fig1}, each surface Dirac cone has one nearby magnetic-doped TI layer, which may be the same layer with the exchange splitting $J_{i\alpha}=J_S$ or the adjacent layer with the exchange splitting $J_{i\alpha}=J_D$.  The Dirac cones localized on different surfaces can be coupled by the Dirac cone hoppings and exchange splittings and thus get renormalized.  Note that the hoppings beyond the nearest-neighbor Dirac cones are neglected because they are minor~\cite{C.Lei} and will not change the main conclusions in this paper.  In our calculations, we take $-\Delta_S=1$ as the unit of energy and set the parameter $\delta=\frac{J_D}{J_S}$. 

To calculate the Chern number in the TI multilayer structures, we follow the previous works~\cite{Y.X.Wang2021, J.Wang} and try to judge the Chern number by exactly capturing the evolution of the phase boundaries with the parameters, as the Chern number change is closely connected to the gap closings.  To help the Chern number judgment, two limiting cases need to be considered: (i) $J_S=0$ of no magnetic-doping case, where the TRS is preserved and the Chern number must be vanishing, with the renormalized Dirac cones always appearing in pairs with opposite masses;  (ii) $\Delta_D=0$ of the limiting isolated-layer case, where the mass of each renormalized Dirac cone $M_i$ can be calculated analytically and then the Chern number is obtained.  As the Dirac cones on all surfaces own the positive chirality $\chi_i=1$ and each Dirac cone can contribute a component $C_i$ to the Chern number, we have 
\begin{align}
C=\sum_i C_i, \quad \text{with} \quad
C_i=\frac{1}{2}\text{sgn}(M_i). 
\end{align}
Compared with the common Chern number calculations by using the Kubo's formula~\cite{D.J.Thouless} or the Fukui's algorithm~\cite{T.Fukui}, where the exact diagonalization is needed to obtain the eigenenergy and eigenstate for each wavevector in the Brillouin zone (BZ), the above method requires much less computational resources and time, and is expected to be extended to the antiferromagnetic configuration as well as other topological systems.  

\begin{figure}
	\includegraphics[width=8.8cm]{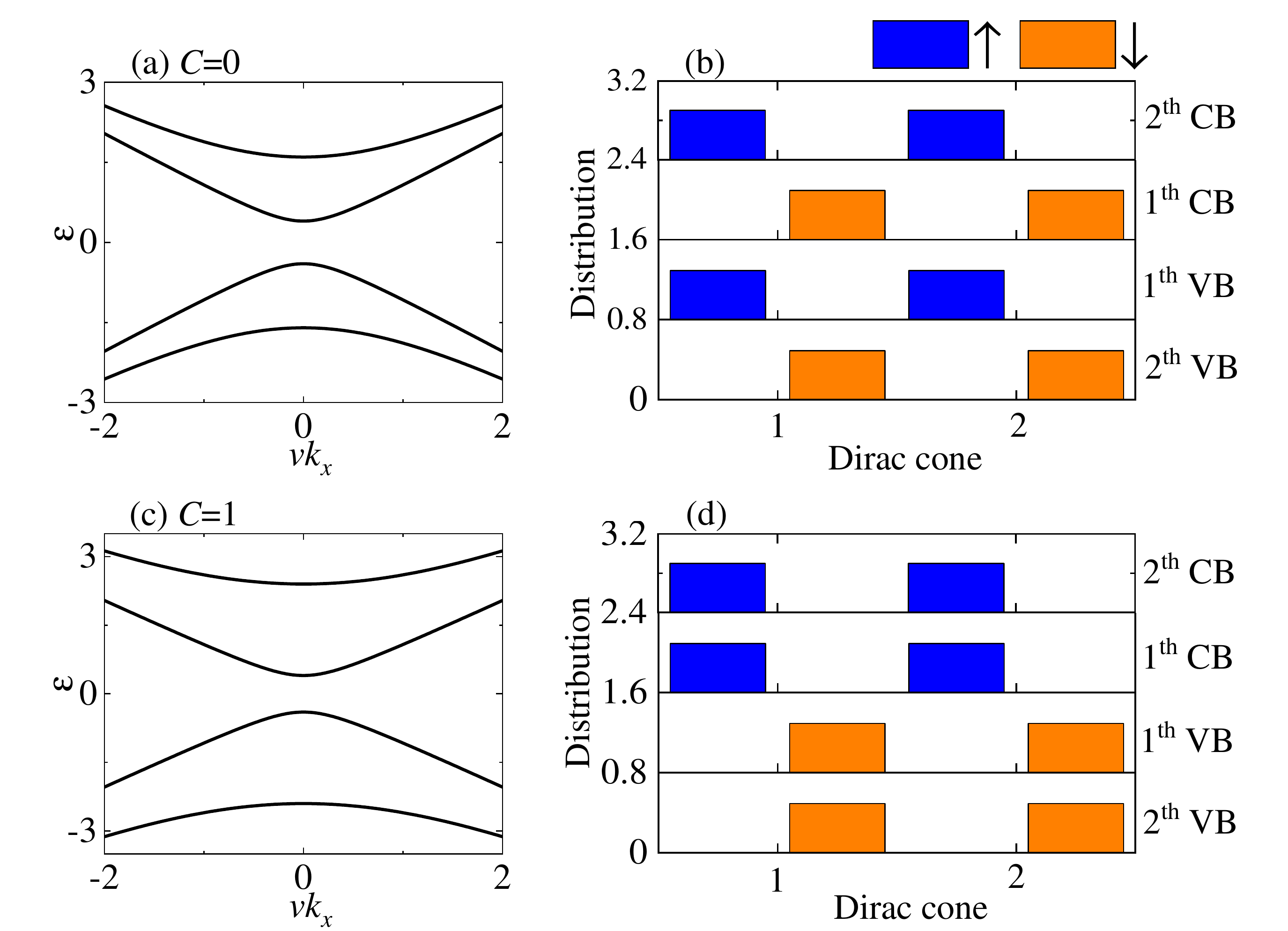}
	\caption{(Color online) The energy bands $\varepsilon\sim v k_x$ when setting $k_y=0$, and the wavefunction distributions at the $\Gamma$ point on each Dirac cone in the magnetic-doped TI thin film.  We choose $J_S=0.4$ in (a)$-$(b), and $J_S=1.6$ in (c)$-$(d).  In the wavefunction distributions, the blue and orange bars denote the upspin and downspin contributions, respectively and the bar heights are proportional to the weights of the distributions.  For clarity, the neighboring distributions are shifted vertically by 0.8. }
	\label{Fig2}	
\end{figure}

\section{magnetic-doped TI thin film} 

First we consider the magnetic-doped TI thin film.  Within the Dirac cone model, a pair of Dirac cones with the same chirality are localized on the top and bottom TI surfaces and there exist the intralayer hoppings between the two Dirac cones.  In the four-orbit basis $\begin{pmatrix}
|1\uparrow\rangle& |1\downarrow\rangle& |2\uparrow\rangle& |2\downarrow\rangle
\end{pmatrix}^T$, where $1/2$ denotes the top/bottom surface state and $\uparrow/\downarrow$ the upspin/downspin state, respectively, the Hamiltonian is~\cite{R.Yu}
\begin{align}
H(\boldsymbol k)=\begin{pmatrix}
J_S& -ivk_-& \Delta_S& 0
\\
ivk_+& -J_S& 0& \Delta_S
\\
\Delta_S& 0& J_S& ivk_-
\\
0& \Delta_S& -ivk_+& -J_S
\end{pmatrix}, 
\label{mTI}
\end{align}
here $k_\pm=k_x\pm ik_y$.  

To see the Dirac cone renormalization, we use the unitary matrix~\cite{R.Yu, Y.X.Wang2018}
\begin{align}
U=\frac{1}{\sqrt2}\begin{pmatrix}
1& 0& 1& 0\\
0& 1& 0& -1\\
1& 0& -1& 0\\
0& -1& 0& -1
\end{pmatrix}. 
\end{align}
Then the Hamiltonian is transformed as
\begin{align}
&H_T(\boldsymbol k)=UH(\boldsymbol k)U^{-1}
\nonumber\\
&=\begin{pmatrix}
J_S-|\Delta_S|& -ivk_-& 0& 0
\\
ivk_+& -J_S+|\Delta_S|& 0& 0
\\
0& 0& J_S+|\Delta_S|& ivk_-
\\
0& 0& -ivk_+& -J_S-|\Delta_S|
\end{pmatrix},
\end{align}
which becomes block diagonalized.  It shows that the chirality of the renormalized Dirac cones in the upper/lower block remains the same as in Eq.~(\ref{mTI}), but the Dirac cone mass in the upper/lower block will get reduced/increased by the intralayer Dirac cone hopping.  So the Chern number of the system is obtained as 
\begin{align}
C=\frac{1}{2}\text{sgn}\big(J_S-|\Delta_S|\big)+\frac{1}{2}\text{sgn}\big(J_S+|\Delta_S|\big).  
\label{s-layer} 
\end{align}
We can see that (i) when $J_S<|\Delta_S|$, the Dirac cone mass is negative in the upper block and positive in the lower block, and thus the Chern number $C=0$; (ii) when $J_S>|\Delta_S|$, the Dirac cone mass in the upper block becomes positive, thus the band inversion occurs and $C=1$.  So in a magnetic-doped TI thin film, the nontrivial $C=1$ phase can be observed only when the magnetic-doping reaches a certain ratio, which is consistent with the experimental observations in the Cr-doped or V-doped (Bi,Sb)$_2$Te$_3$ thin film~\cite{C.Z.Chang2013, X.Kou2014, Y.Feng, J.G.Checkelsky, C.Z.Chang2015}.  Note that the Dirac cone model cannot support the high-$C$ phase in the TI thin film, which, although predicted in theory~\cite{H.Jiang, J.Wang, Y.X.Wang2021}, but to our knowledge, has not yet been reported in experiment.  

\begin{figure*}
	\includegraphics[width=18.4cm]{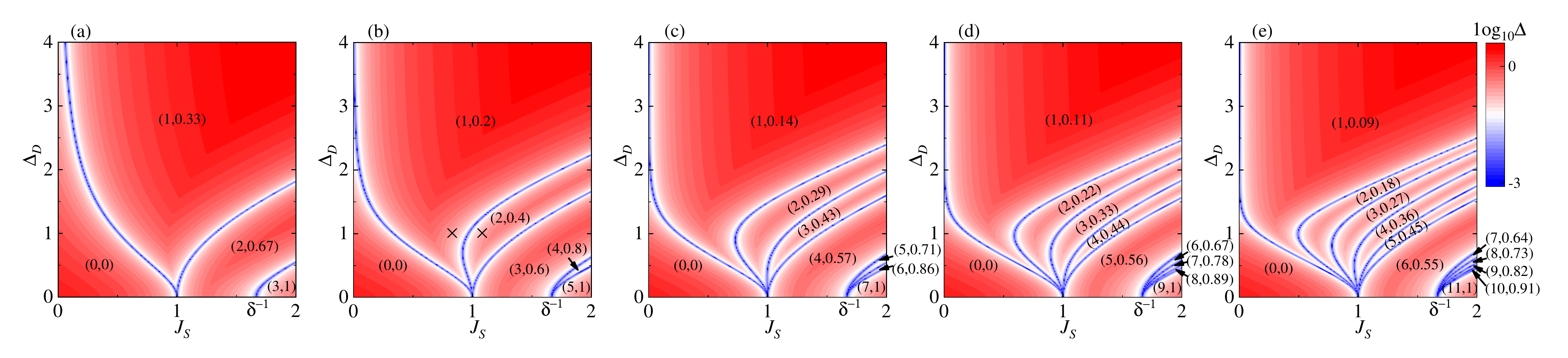}
	\caption{(Color online) Phase diagrams of the TI multilayer structures in the parameter space $(J_S,\Delta_D)$, with the Chern number $C$ and the spin polarization $\eta$ at the $\Gamma$ point being labeled as $(C,\eta)$.  The contour scale represents the magnitude of log$_{10}\Delta$, where $\Delta$ is the energy gap of the lowest bands, and the bright blue lines denote that the gap is closed.  We choose the parameter $\delta=0.6$, the layer number $N_2=1-5$ in (a) to (e) and $N_1=N_2+1$. }
	\label{Fig3}
\end{figure*}

In Fig.~\ref{Fig2}, we plot the energy bands and the wavefunction distributions at the $\Gamma$ point in the $C=0$ and $C=1$ phases.  For the Hamiltonian including only the linear $k$ terms, the normal and inverted bands are identical [Figs.~\ref{Fig2}(a) and (c)], so there is no signature of the band inversions in the band structures.  If the second order $k$ term $(\propto k^2)$ is included, the band inversion can be found as an additional energy extremum at the nonzero wave vector, whereas a single extremum at the $\Gamma$ point remains in the normal band~\cite{R.Yu}.  On the other hand, the band inversions can be manifested in the wavefunction distributions at the $\Gamma$ point.  In the $C=0$ phase, the orbital contributions to the first valence band (VB) are $\frac{1}{\sqrt2}(-|1\uparrow\rangle-|2\uparrow\rangle)$, and those to the first conduction band (CB) are $\frac{1}{\sqrt2}(-|1\downarrow\rangle+|2\downarrow\rangle)$ [Fig.~\ref{Fig2}(b)], while in the $C=1$ phase, the contributions to the first VB and first CB are interchanged [Fig.~\ref{Fig2}(d)], meaning that all VBs at the $\Gamma$ point are downspin polarized.

\section{TI multilayer structures} 

Next, for the TI multilayer structures, the Chern number phase diagrams are plotted in the parametric space $(J_S,\Delta_D)$ in Fig.~\ref{Fig3}, with the layer number $N_2$ increasing from 1 to 5 in (a) to (e) and $N_1=N_2+1$.  Experimentally, tuning the parameters $J_S$ and $\Delta_D$ is quite feasible, as the exchange splitting $J_S$ is expected to increase with the magnetic-doping or increase with the decreasing temperature, and the interlayer hopping $\Delta_D$ is expected to increase when the van der Waals gap is narrowed by the external pressure~\cite{C.Lei, W.T.Guo}.  In the phase diagrams, the contour scale represents the magnitude of log$_{10}\Delta$, where $\Delta$ is the energy gap of the lowest bands and the bright blue lines denote that the gap is closed.  

Fig.~\ref{Fig3} shows that the Chern number $C$ can increase from zero up to its highest value $C=N_1+N_2$.  When $J_S=0$ of no magnetic-doping case, the Chern number always vanishes due to the presence of the TRS.  On the other hand, when $\Delta_D=0$ of the isolated-layer case, similar to Eq.~(\ref{s-layer}), the Chern number of the TI multilayer structures is obtained as 
\begin{align}
C=&\frac{N_1}{2}\Big[\text{sgn}\big(J_S-|\Delta_S|\big)+\text{sgn}\big(J_S+|\Delta_S|\big)\Big]
\nonumber\\
&+\frac{N_2}{2}\Big[\text{sgn}\big(J_D-|\Delta_S|\big)+\text{sgn}\big(J_D+|\Delta_S|\big)\Big].  
\label{m-layer}
\end{align}
We can see that when the exchange splitting $J_S$ increases, the negative mass $J_S-|\Delta_S|$ of the Dirac cone in the magnetic-doped TI layer becomes positive at the critical point $J_S=|\Delta_S|$, which is $N_1$-fold degenerate; whereas the negative mass $J_D-|\Delta_S|$ of the Dirac cone in the magnetic-undoped TI layer becomes positive at the critical point $J_D=|\Delta_S|$, which is $N_2$-fold degenerate.  So the Chern number evolutions with $J_S$ are as follows: (i) when $J_S<|\Delta_S|$, the Chern number vanishes $C=0$;  (ii) when $|\Delta_S|<J_S<\frac{|\Delta_S|}{\delta}$, $C=N_1$; (iii) when $J_S>\frac{|\Delta_S|}{\delta}$, the Chern number increases to its highest value $C=N_1+N_2$, where the band inversions occur for all negative-mass Dirac cones. 

\begin{figure}
	\includegraphics[width=8.8cm]{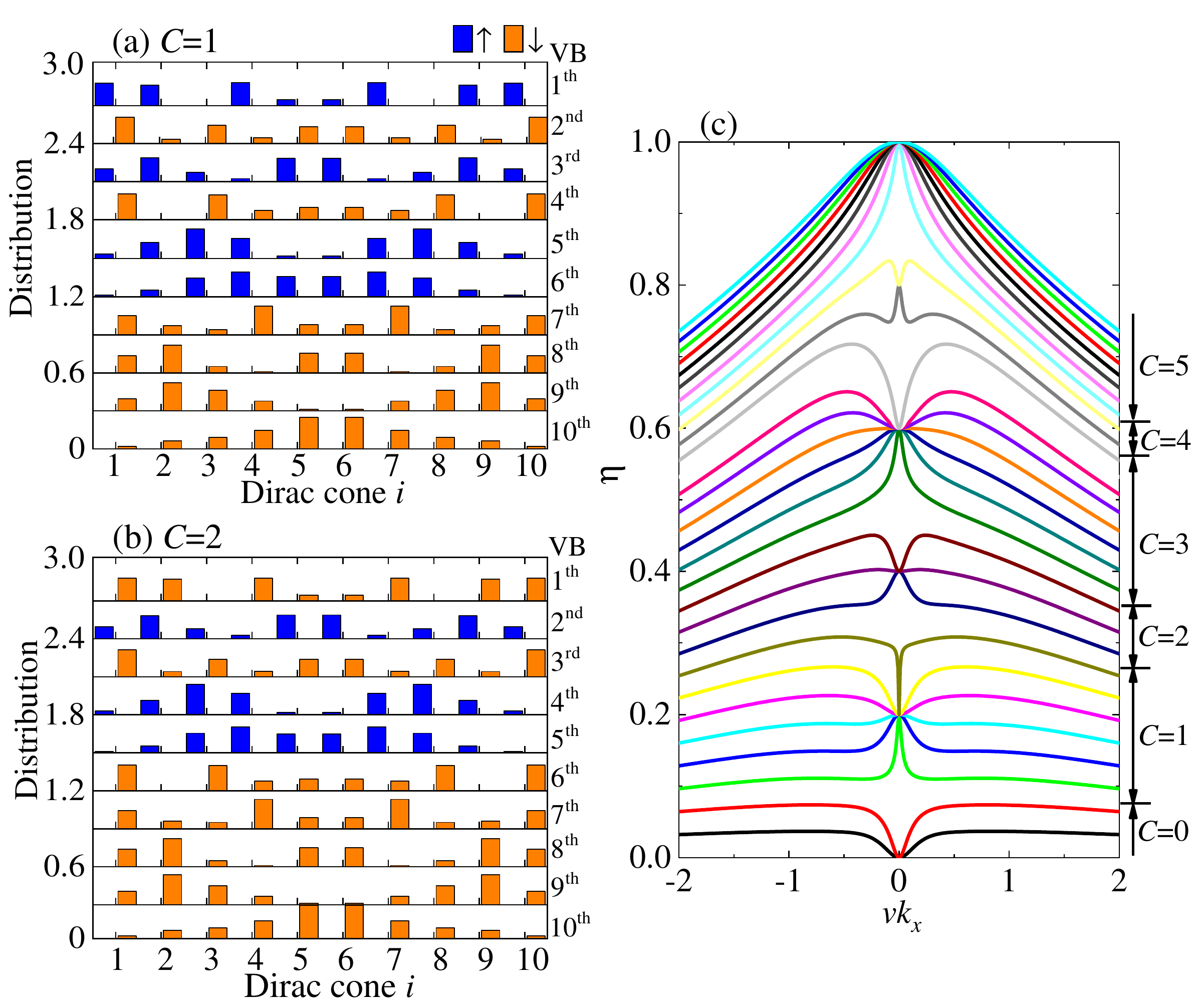}
	\caption{(Color online) (a)-(b) The wavefunction distributions of all VBs at the $\Gamma$ point on each Dirac cone in the TI multilayer structures.  The blue and orange bars denote the upspin and downspin contributions, respectively, and the bar heights are proportional to the weights of the distributions.  For clarity, the neighboring distributions are shifted vertically by 0.3.  (c) The spin polarization versus $vk_x$, where $J_S$ increases from 0.12 to 3.6 and the corresponding Chern number is labeled in the right.  We choose $N_2=2$ and $N_1=N_2+1$, $\Delta_D=1$, $J_S=0.84$ in (a), $J_S=1.08$ in (b), and $k_y=0$ in (c). }
	\label{Fig4}
\end{figure}

When the interlayer hopping $\Delta_D$ is nonvanishing, the Dirac cones in different TI layers will get coupled and further renormalized, which thus breaks the Dirac cone degeneracy.  The renormalized Dirac cones may not be localized in one TI layer, but extended to all layers [see Figs.~\ref{Fig4}(a) and (b)].  As a result, with the increasing of $J_S$, the band inversion occurs successively for the renormalized Dirac cones.  Each time a phase boundary is crossed, the band inversion occurs for one negative-mass Dirac cone and the Chern number will be changed by one.  When $\Delta_D$ increases, we can see that all phase boundaries originating from $J_S=\frac{|\Delta_S|}{\delta}$ move to larger $J_S$; on the other hand, for the phase boundaries from $J_S=|\Delta_S|$, only one will move to $J_S=0$, meaning that the region spanned by the $C=0$ phase is shrinking so that a minor magnetic doping can drive the transition from the $C=0$ phase to $C=1$, while the remaining $N_1-1$ ones all move to larger $J_S$.

In the phase diagrams, if the magnetic-doping is small, the value of the critical point $J_S=\frac{|\Delta_S|}{\delta}$ at $\Delta_D=0$ will become larger; while if the magnetic-ordered septuple layer is formed, such as in MnBi$_2$Te$_4$~\cite{M.M.Otrokov}, where the exchange splitting is greatly strengthened, $\delta$ is expected be close to 1 and thus the critical point is close to $J_S=|\Delta_S|$.  In the experiment~\cite{Y.F.Zhao}, the thicknesses of the magnetic-doped and undoped layers are taken as $d=3$ nm and $d=4$ nm, respectively.  The different TI layer thicknesses and properties can lead to the unequal $\Delta_S$ in the different layers, which, however, will not change the basic physics and the structure of the phase diagrams (see Appendix). 

Next, we study the wavefunction distribution and the spin polarization in the TI multilayer structures.  The spin polarization $\eta_{\boldsymbol k}$ is defined as 
\begin{align}
\eta_{\boldsymbol k}=
\frac{\sum_{n\in\text{occ}}\big(|\psi_{n,\downarrow}(\boldsymbol k)|^2
-|\psi_{n,\uparrow}(\boldsymbol k)|^2\big)}
{\sum_{n\in\text{occ}}\big(|\psi_{n,\downarrow}(\boldsymbol k)|^2
+|\psi_{n,\uparrow}(\boldsymbol k)|^2\big)}, 
\end{align}
where $\psi_{n,\uparrow/\downarrow}$ is the upspin/downspin component of the wavefunction with the band index $n$, and $n$ is to be summed over all occupied states.  

Because the band inversion is accompanied by the change of the orbital contribution at the $\Gamma$ point from upspin to downspin, so $\eta_{\Gamma}$ is directly related to the Chern number,  
\begin{align}
\eta_{\Gamma}=\frac{C}{N_1+N_2}.  
\label{polarization}
\end{align}
The above relation can also be demonstrated by the numerical results in Fig.~\ref{Fig3}.  For example, in Fig.~\ref{Fig4}(a) and (b), with the parameters chosen as the points labeled by the crosses in Fig.~\ref{Fig3}(b), we plot the wavefunction distributions of all VBs at the $\Gamma$ point on each Dirac cone.  For Chern number phase changing from $C=1$ to $C=2$, we observe that besides some band movements, the orbital contribution to the first VB changes from upspin to downspin, and thus the spin polarization $\eta_{\Gamma}$ will increase by $\frac{1}{N_1+N_2}$.  At the highest Chern number phase $C=N_1+N_2$, the spin polarization can reach its saturation value, $\eta_{\Gamma}=1$.

We also investigate the spin polarization at nonzero wave vectors.  In Fig.~\ref{Fig4}(c), when setting $k_y=0$, $\eta$ is plotted as a function of $vk_x$, where the different lines correspond to $J_S$ increasing from $0.12$ to $3.6$.  We can see that although $\eta_{\boldsymbol k}$ may increase with $|k_x|$ around the $\Gamma$ point, it decreases when $|k_x|$ is large enough.  On the other hand, $\eta_{\boldsymbol k}$ always increases with $J_S$.  So at the asymptotically strong Zeeman splitting, the system may be driven into the ferromagnetic Chern insulator phase~\cite{W.Wang}.  In the experiment, the spin polarization can be detected by the spin-polarized angle-resolved photoemission spectroscopy~\cite{J.A.Sobota}.

\begin{figure}
	\includegraphics[width=9cm]{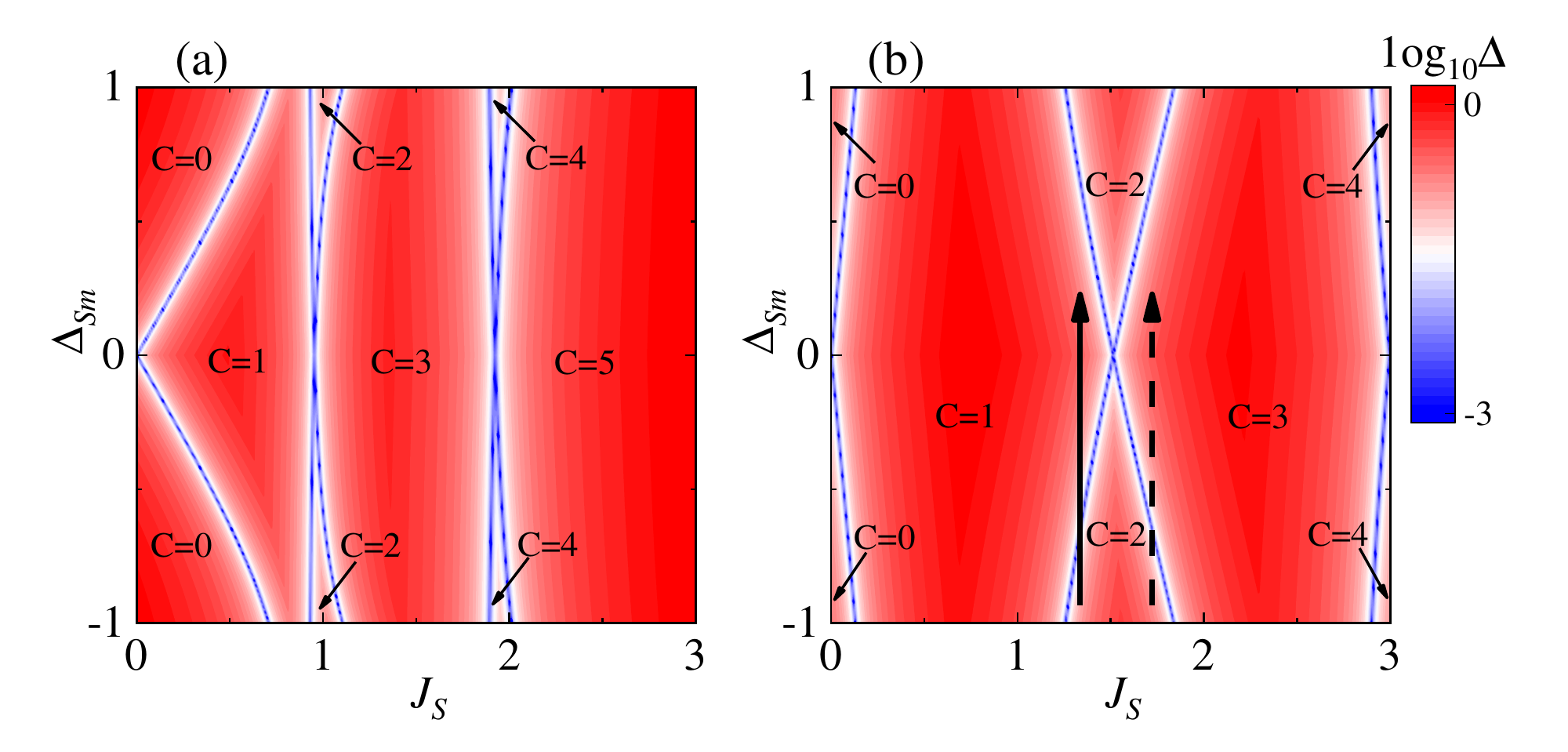}
	\caption{(Color online) Phase diagrams of the TI multilayer structures in the parameter space of $(J_S,\Delta_{Sm})$, with the Chern number $C$ being labeled.  The contour scale represents the magnitude of log$_{10}\Delta$, where $\Delta$ is the energy gap of the lowest bands.  The bright blue lines denote that the gap is closed.  We choose the parameter $\delta=0.6$, the layer number $N_2=2$, $N_1=N_2+1$, and $\Delta_D=0.5$ in (a) and $\Delta_D=1.5$ in (b).}
	\label{Fig5}
\end{figure}

The recent experiment that implemented on the TI multilayers [(Bi,Sb)$_{2-x}$Cr$_x$Te$_3-$(Bi,Sb)$_2$Te$_3$]$_{N_2}-$(Bi,Sb)$_{2-x}$Cr$_x$Te$_3$ reported the observations of the high-$C$ phase with $C=N_2$ when the magnetic-doping ratio is $x=0.24$~\cite{Y.F.Zhao}.  According to the above analysis, this suggests that at such a ratio, the band inversion only occurs in $N_2$ negative-mass Dirac cones.  When $\Delta_D=0$, all these Dirac cones are localized in the magnetic-doped TI layers.  In their work, the authors also reported that for the TI multilayers $N_1=3$ and $N_2=2$, the Chern number can change from $C=2$ to $C=1$ when the middle magnetic-doped layer thickness was tuned to decrease from $d=4$ nm to zero~\cite{Y.F.Zhao}.  They suggested that when the magnetic-doped layer thickness decreases, a pair of nontrivial interface states will disappear, so the Chern number is reduced by one.  Here in the Dirac cone model, this observation could be instead explained by assuming that the intralayer hopping within the middle magnetic-doped TI layer $\Delta_{Sm}$ will change from negative to positive when the layer thickness decreases.  This means that when the layer is thick enough, the Dirac cones attract each other so that $\Delta_{Sm}$ is negative, while when the layer thickness decreases and becomes small, the Dirac cones become less attractive and even repel each other when a critical thickness is acrossed, so $\Delta_{Sm}$ becomes positive.  In our calculations, the chosen model parameters $\Delta_S=-1$ and $\Delta_D>0$ are consistent with the above assumptions.  Moreover, in Ref.~\cite{C.Lei}, by comparing the model parameters with the DFT band calculations, the authors obtained the opposite $\Delta_S$ and $\Delta_D$, which also supports our assumptions. 

To see the effect of $\Delta_{Sm}$ on the Chern number modulation, in Fig.~\ref{Fig5}, the phase diagrams of the TI multilayer structures are plotted in the parameter space $(J_S,\Delta_{Sm})$, with $\Delta_D=0.5$ in (a) and $\Delta_D=1.5$ in (b).  We can see that in Fig.~\ref{Fig5}(a), there is no $C=2$ to $C=1$ phase transition along the vertical direction.  However, in Fig.~\ref{Fig5}(b), the $C=2$ to $C=1$ phase transition can be captured along the left solid arrow, which agrees with the experiment~\cite{Y.F.Zhao}.  More interestingly, in Fig.~\ref{Fig5}(b), when the magnetic-doping increases and the exchange splitting lies within the range $1.51<J_S<1.84$, there can exist the $C=2$ to $C=3$ phase transition, \textit{e.g.}, along the right dashed arrow.  Therefore, as long as the layer thickness is not vanishing and the Dirac cones are still present on the surfaces, the Chern number may get increased.  The conclusion is quite different from the interface Dirac state mechanism, but also roots in the band inversion of the Dirac cone.

\section{other TI multilayer structures}

\begin{figure}
	\includegraphics[width=9cm]{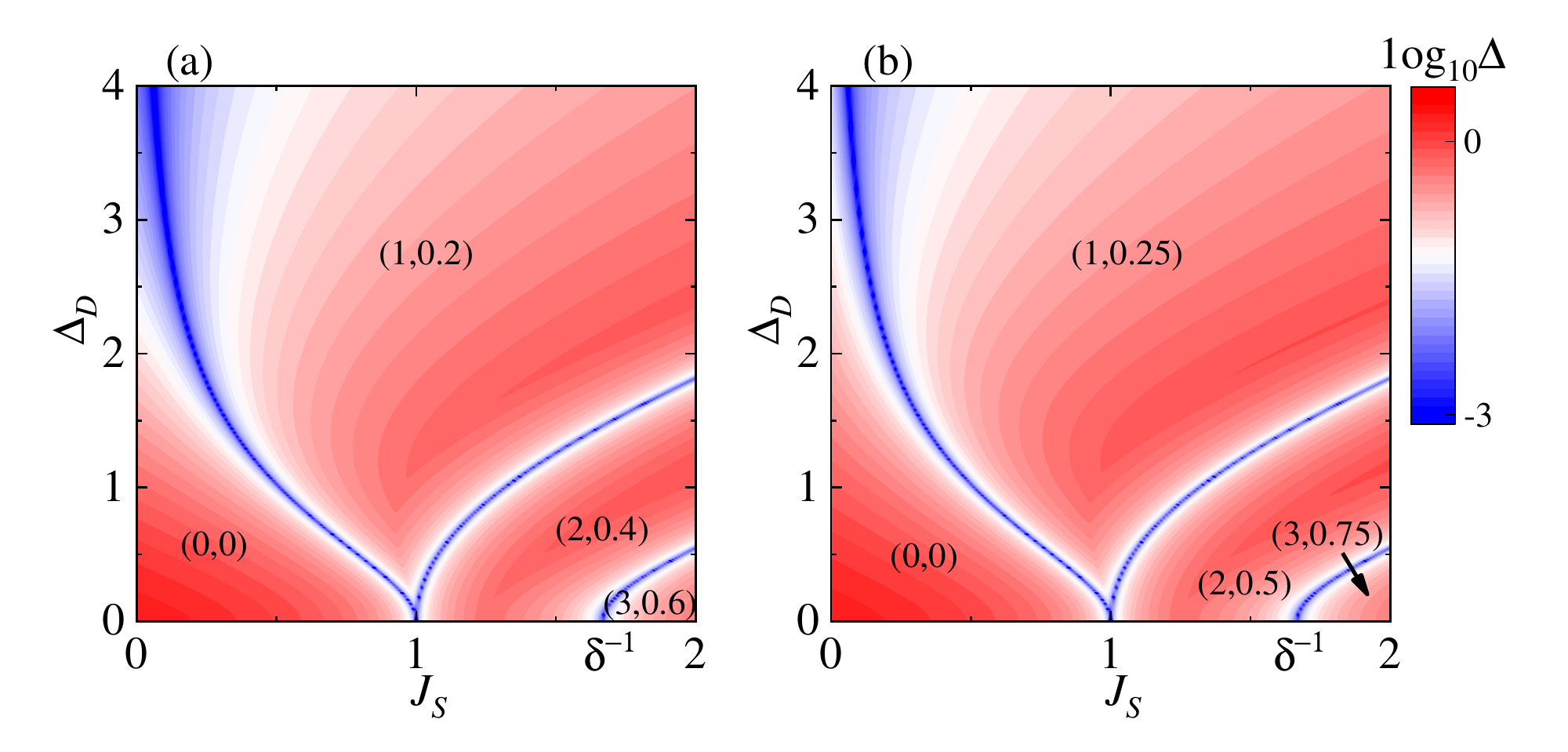}
	\caption{(Color online) Phase diagrams of two alternative TI multilayer structures in the parameter space of $(J_S,\Delta_D)$, with the Chern number $C$ and the spin polarization $\eta$ at the $\Gamma$ point being labeled as $(C,\eta)$.  The contour scale represents the magnitude of log$_{10}\Delta$, where $\Delta$ is the energy gap of the lowest bands and the bright blue lines denote that the gap is closed.  We choose the parameter $\delta=0.6$, the layer number $N_1=2$, $N_2=N_1+1$ in (a) and $N_1=N_2=2$ in (b).}
	\label{Fig6}
\end{figure}

To make comparisons, we study another two TI multilayer structures, with the Chern number phase diagrams plotted in Fig.~\ref{Fig6}.  In Fig.~\ref{Fig6}(a), we choose the layer number $N_1=2$ and $N_2=3$ so that the magnetic-doped TI layers are sandwiched between the undoped ones, whereas in Fig.~\ref{Fig6}(b), we choose $N_1=N_2=2$ and the number of the magnetic-doped TI layer is equivalent to the undoped layer.

In the isolated layer case of $\Delta_D=0$, for the topmost undoped TI layer, the masses of the two renormalized Dirac cones are calculated as 
\begin{align}
M_{t,\pm}^o=&\frac{1}{2}J_D\pm\frac{1}{2}\sqrt{J_D^2+4\Delta_S^2}. 
\end{align}
Clearly, $M_{t,\pm}^o$ is always positive/negative and there is no sign change.  So the contributions of the two Dirac cones to the Chern number are always zero.  Cases are the same for the bottommost undoped TI layer, if exists.  This means that the Chern number is contributed by the magnetic-doped TI layers as well as the sandwiched undoped TI layers, but not the topmost or bottommost undoped TI layer.  This conclusion also holds when $\Delta_D$ is nonvanishing.  Therefore, as shown in Figs.~\ref{Fig6}(a) and (b), the highest Chern number can only reach $C=2N_1-1$. 

For the spin polarization $\eta$ at the $\Gamma$ point, the results are shown in Fig.~\ref{Fig6}.  We can see that the relation in Eq.~(\ref{polarization}) of the spin polarization to the Chern number still holds.  Comparing Figs.~\ref{Fig6}(a) and (b) with Fig.~\ref{Fig3}(a), we find that although the nontrivial Chern number phases span the similar regions, the spin polarizations are quite different.  As analyzed above, even when the Chern number takes its highest value, the spin polarization at the $\Gamma$ point cannot be saturated in the two TI multilayer structures.  

We note that in Fig.~\ref{Fig6}, the gaps are well opened in the Chern insulator phases.  Because the Fermi energy needs to be tuned in the gap, so the large gap favors the experimental observations.  This conclusion is quite different from our previous work~\cite{Y.X.Wang2021}, where the gaps of the Chern insulator phases were found to be too small to hinder the experimental observations.

\section{TI superlattice structure} 

In addition, we study the TI superlattice structure with the periodic boundary conditions in the $z$ direction.  Note that in a unit cell, the topmost and bottommost Dirac cones may have two nearby  magnetic-doped TI layers.  As more and more magnetic-ordered TIs have been successfully fabricated in experiments~\cite{M.M.Otrokov, J.Wu, R.C.Vidal, I.I.Klimovskikh}, the TI superlattice considered here may also be realized in the future.

\begin{figure}
	\includegraphics[width=9.4cm]{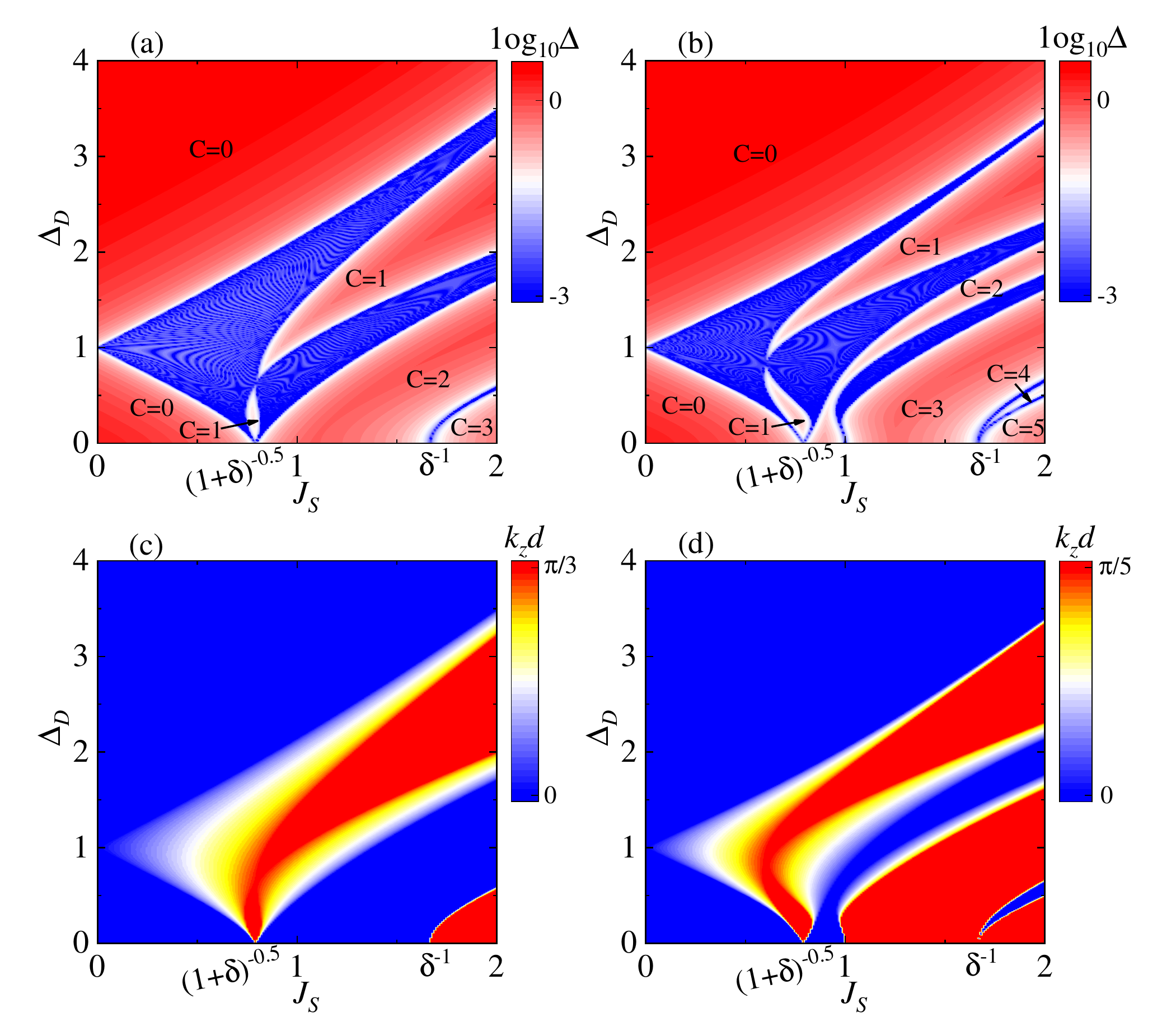}
	\caption{(Color online) Phase diagrams of the TI superlattice structures in the parameter space of $(J_S,\Delta_D)$, and the corresponding $k_z$ value for the energy gap.  In (a) and (b), the Chern number $C$ is labeled and the contour scale represents the magnitude of log$_{10}\Delta$, where $\Delta$ is the energy gap of the lowest bands.  The bright blue regions denote that the gap is closed, and the system lies in the gapless WSM phase.  We choose the parameter $\delta=0.6$, the layer number $N_2=1$ in (a) and (c), $N_2=2$ in (b) and (d), and $N_1=N_2+1$. }
	\label{Fig7}
\end{figure} 

The phase diagrams for the TI superlattice are plotted in Figs.~\ref{Fig7}(a) and (b), where the unit cell includes the same TI layer number as in Figs.~\ref{Fig3}(a) and (b), respectively.  The bright blue regions denote that the energy gaps are closed so that the system lies in the gapless WSM phase.  Figs.~\ref{Fig7}(a) and (b) show that the WSM phases can span large parameter regions and separate the distinct Chern insulator phases.  This is consistent with the previous studies~\cite{A.A.Burkov, C.Lei}, but is quite different from the TI multilayer structures, where no WSM phase exists in Fig.~\ref{Fig3}.  This is explained that the wave vector $k_z$ needs to take a specific value to close the gap so that the WSM phase can be accommodated in the TI superlattice~\cite{A.A.Burkov, A.A.Zyuzin}, but $k_z$ cannot be modulated in the TI multilayer structures.  We find that the wave vector $k_z$ for the gap can also be used to distinguish the Chern insulator and WSM phases, where in the former, $k_z$ takes the fixed value $0$ or $\frac{\pi}{(N_1+N_2)d}$ that situates in the BZ center or boundary, while in latter, $k_z$ gradually changes from $0$ to $\frac{\pi}{(N_1+N_2)d}$.  Note that in the superlattice structure, due to the enlarged unit cell, the BZ in the $z$ direction has been folded into $[-\frac{\pi}{(N_1+N_2)d},\frac{\pi}{(N_1+N_2)d}]$.   

When $J_S=0$ of no magnetic-doping case, the masses of the $2(N_1+N_2)$ renormalized Dirac cones can be calculated analytically as 
\begin{align} 
M_{\pm}^s=\pm\sqrt{\Delta_S^2+\Delta_D^2+2\Delta_S\Delta_D
\text{cos}\big(k_z d+\frac{2\pi\alpha}{N_1+N_2}\big)}, 
\end{align}
with the layer index $\alpha=1,2,\cdots,N_1+N_2$.  The above equation shows that the masses always appear in pairs.  One can check that the gap can get closed only when $M_{N_1+N_2,\pm}^s$ vanish, with the conditions of $k_z=0$ and $\Delta_D=|\Delta_S|$, as shown in Fig.~\ref{Fig7}. 

On the other hand, when $\Delta_D=0$ of isolated-layer case, the masses of the two renormalized Dirac cones on the topmost magnetic-doped TI layer in a unit cell are obtained as
\begin{align}
M_{t,\pm}^s=J_S+\frac{1}{2}J_D\pm\frac{1}{2}\sqrt{J_D^2+4\Delta_S^2}, 
\label{msuper}
\end{align}
which are the same as the bottommost layer in a unit cell.  One can infer from Eq.~(\ref{msuper}) that the mass $M^s_{t,-}$ changes from negative to positive at the critical point $J_S=\frac{|\Delta_S|}{\sqrt{1+\delta}}$ and is twofold degenerate.  The other critical points at $\Delta_D=0$ are the same as those in TI multilayer structures, suggesting that the Chern number behaviors are similar.

\section{Discussions and Conclusions}

In summary, we have studied the Chern number phase diagrams in the TI multilayer structures by using the Dirac cone model, where the Chern number is effectively calculated by capturing the evolutions of the phase boundaries with the parameters.  We find that the magnetic doping can drive the band inversion of the Dirac cones and thus induce the high-$C$ phase transitions.  In the highest Chern number phase, the occupied states are downspin polarized at the $\Gamma$ point.

Compared with the previous theoretical works that used the plane wave expansion method on the $\boldsymbol k\cdot \boldsymbol p$ model~\cite{Y.F.Zhao, Y.X.Wang2021}, our results show the following differences: (i) the inverted bands are related to the renormalized Dirac cones; (ii) the Chern number may get increased even when the thickness of the middle magnetic-doped TI layer decreases; (iii) the gaps are well opened in the Chern insulator phase in another two TI multilayer structures.  The differences are attributed to the fact that the surface states of each TI layer in the multilayer structures are well captured by the Dirac cone model, but may not be correctly described by the plane-wave expansion method.  We hope that these conclusions, especially (ii) and (iii), would be demonstrated in the future TI multilayer experiments.

Besides the magnetic-doped TI multilayers, the magnetic-ordered ones, as well as the factors that cannot be included in the DFT calculations, such as disorder and magnetic field, can also be described by the Dirac cone model.  We expect that the Dirac cone model to be widely used to explain more emergent phenomena in the TI multilayer structures, such as the axion insulator~\cite{M.Mogi, D.Xiao}.

\section{Acknowledgments} 

This work was supported by the National Natural Science Foundation of China (Grant No. 11804122 and No. 11905054), the China Postdoctoral Science Foundation (Grant No. 2021M690970), and the Fundamental Research Funds for the Central Universities of China.

\section{Appendix: unequal intralayer Dirac cone hoppings}

\begin{figure}
	\includegraphics[width=9cm]{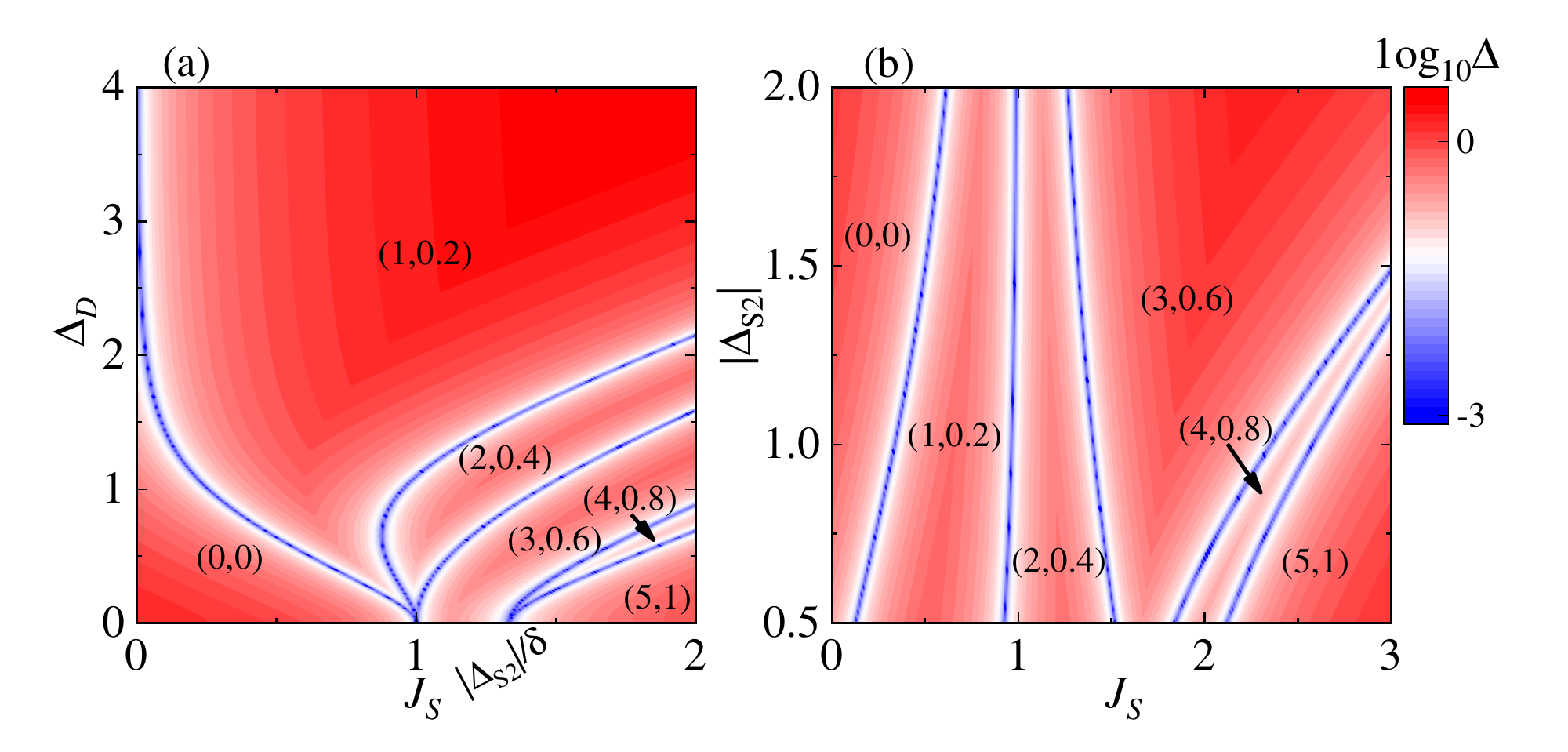}
	\caption{(Color online)  Phase diagrams of the TI multilayer structures in the parameter space $(J_S,\Delta_D)$ in (a) and $(J_S,|\Delta_{S2}|)$ in (b), with the Chern number $C$ and the spin polarization $\eta$ at the $\Gamma$ point being labeled as $(C,\eta)$.  The contour scale represents the magnitude of log$_{10}\Delta$, where $\Delta$ is the energy gap of the lowest bands, and the bright blue lines denote that the gap is closed.  We choose the parameter $\delta=0.6$, the layer number $N_2=2$, $N_1=N_2+1$, $\Delta_{S1}=-1$, and in (a) $\Delta_{S2}=-0.8$, in (b) $\Delta_D=0.9$.}
	\label{Fig8}
\end{figure}

Here we discuss the effect of the unequal intralayer Dirac cone hoppings on the phase diagrams. 
We use $\Delta_{S1/2}$ to represent the intralayer Dirac cone hopping in the magnetic-doped/updoped TI layer. 

When $\Delta_D=0$, the Chern number in Eq.~(\ref{m-layer}) becomes
\begin{align}
C=&\frac{N_1}{2}\Big[\text{sgn}\big(J_S-|\Delta_{S1}|\big)+\text{sgn}\big(J_S+|\Delta_{S1}|\big)\Big]
\nonumber\\
&+\frac{N_2}{2}\Big[\text{sgn}\big(J_D-|\Delta_{S2}|\big)+\text{sgn}\big(J_D+|\Delta_{S2}|\big)\Big]. \end{align}
The above Chern number expression shows that the critical points are located at $J_S=|\Delta_{S1}|$ and $J_S=\frac{|\Delta_{S2}|}{\delta}$, as shown in Fig.~\ref{Fig8}(a).  More importantly, the structure of the phase diagram in Fig.~\ref{Fig8}(a) is similar with Fig.~\ref{Fig3}(b).  When setting $\Delta_D=0.9$, in Fig.~\ref{Fig8}(b), we plot the phase diagram in the parametric space  $(J_S,|\Delta_{S2}|)$.  We observe that as $|\Delta_{S2}|$ increases, the left three phase boundaries show minor changes, while the right two phase boundaries will move to higher $J_S$.  So for the unequal $\Delta_{S1}$ and $\Delta_{S2}$, the qualitative results of the phase diagram remain unchanged.


\begin{references} 

\bibitem{D.J.Thouless} 
Thouless D J, Kohmoto M, Nightingale M P, and Nijs M den, 
1982 Phys. Rev. Lett. {\bf49} 405

\bibitem{F.D.M.Haldane}
Haldane F D M, 
1988 Phys. Rev. Lett. {\bf61} 2015

\bibitem{X.L.Qi} 
Qi X L, Hughes T L, and Zhang S C, 
2008 Phys. Rev. B {\bf78} 195424

\bibitem{C.Z.Chang2013}
Chang C Z, Zhang J, Feng X, Shen J, Zhang Z, Guo M, Li K, Ou Y, Wei P, Wang L L, Ji Z Q, Feng Y, Ji S, Chen X, Jia J, Dai X, Fang Z, Zhang S C, He K, Wang Y, Lu L, Ma X C, and Xue Q K, 
2013 Science {\bf340} 167

\bibitem{R.Yu}
Yu R, Zhang W, Zhang H J, Zhang S C, Dai X, Fang Z,
2010 Science {\bf329} 61

\bibitem{Y.L.Chen}
Chen Y L, Chu J H, Analytis J G, Liu Z K, Igarashi K, Kuo H H, Qi X L, Mo S K, Moore R G, Lu D H, Hashimoto M, Sasagawa T, Zhang S C, Fisher I R, Hussain Z, and Shen Z X, 
2010 Science {\bf329} 659

\bibitem{J.Zhang}
Zhang J, Zhao B, Zhou T, and Yang Z, 
2016 Chin. Phys. B {\bf25} 117308

\bibitem{X.Kou2015}
Kou X, Pan L, Wang J, Fan Y, Choi E S, Lee W L, Nie T, Murata K, Shao Q, Zhang S C, and Wang K L,
2015 Nat. Commun. {\bf6} 8474

\bibitem{J.G.Checkelsky} 
Checkelsky J G, Yoshimi R, Tsukazaki A, Takahashi K S, Kozuka Y, Falson J,  Kawasaki M, and Tokura Y, 
2014 Nat. Phys. {\bf10} 731

\bibitem{Y.Feng}
Feng Y, Feng X, Ou Y, Wang J, Liu C, Zhang L, Zhao D, Jiang T, Zhang S C, He K,  Ma X, Xue Q K, and Wang Y, 
2015 Phys. Rev. Lett. {\bf115} 126801

\bibitem{X.Kou2014}
Kou X, Guo S T, Fan Y, Pan L, Lang M, Jiang Y, Shao Q, Nie T, Murata K, Tang J,  Wang Y, He L, Lee T K, Lee W L, and Wang K L, 
2014 Phys. Rev. Lett. {\bf113} 137201

\bibitem{C.Z.Chang2015}
Chang C Z, Zhao W, Kim D Y, Zhang H, Assaf B A, Heiman D, Zhang S C, Liu C, Chan M H W, and Moodera J S, 
2015 Nat. Mater. {\bf14} 473

\bibitem{Rienks}
Rienks E D L, Wimmer S, S\'anchez-Barriga J, Caha O, Mandal P S, Ruzicka J, Ney A, Steiner H, Volobuev V V, Groiss H, Albu M, Kothleitner G, Michalicka J, Khan S A,  Min\'ar J, Ebert H, Bauer G, Freyse F, Varykhalov A, Rader O and Springholz G, 
2019 Nature {\bf576} 423

\bibitem{H.Deng}
Deng H, Chen Z, Wolo\'s A, Konczykowski M, Sobczak K, Sitnicka J, Fedorchenko I V, Borysiuk J, Heider T, Pluci\'nski L, Park K, Georgescu A B, Cano J and  Krusin-Elbaum L, 
2021 Nat. Phys. {\bf17} 36

\bibitem{Y.F.Zhao}
Zhao Y F, Zhang R, Mei R, Zhou L, Yi H, Zhang Y Q, Yu J, Xiao R, Wang K, Samarth N, Chan M H W, Liu C X, and Chang C Z, 
2020 Nature (London) {\bf588} 419

\bibitem{Y.X.Wang2021}
Wang Y X and Li F, 
2021 Phys. Rev. B {\bf104} 035202

\bibitem{H.Zhang}
Zhang H, Liu C X, Qi X L, Dai X, Fang Z, and Zhang S C,
2009 Nat. Phys. {\bf5} 438

\bibitem{C.X.Liu}
Liu C X, Qi X L, Zhang H J, Dai X, Fang Z, and Zhang S C, 
2010 Phys. Rev. B {\bf82} 045122

\bibitem{J.Wang}
Wang J, Lian B, Zhang H, Xu Y, and Zhang S C,
2013 Phys. Rev. Lett. {\bf111} 136801

\bibitem{B.Zhou}
Zhou B, Lu H Z, Chu R L, Shen S Q, and Niu Q, 
2008 Phys. Rev. Lett. {\bf101} 246807

\bibitem{A.A.Burkov}
Burkov A A and Balents L, 
2011 Phys. Rev. Lett. {\bf107} 127205

\bibitem{A.A.Zyuzin}
Zyuzin A A, Wu S, and Burkov A A, 
2012 Phys. Rev. B {\bf85} 165110

\bibitem{C.Lei}
Lei C, Chen S, and MacDonald A H, 
2020 Proc. Natl. Acad. Sci. USA {\bf117} 27224

\bibitem{T.Fukui}
Fukui T, Hatsugai Y, and Suzuki H, 
2005 J. Phys. Soc. Jpn. {\bf74} 1674

\bibitem{Y.X.Wang2018}
Wang Y X and Li F, 
2018 EPL {\bf123} 37001.

\bibitem{H.Jiang}
Jiang H, Qiao Z, Liu H, and Niu Q, 
2012 Phys. Rev. B {\bf85} 045445

\bibitem{W.T.Guo}
Guo W T, Huang L, Yang Y, Huang Z and Zhang J M, 
2021 New J. Phys. {\bf23} 083030

\bibitem{M.M.Otrokov}
Otrokov M M, Klimovskikh I I, Bentmann H, Estyunin D, Zeugner A, Aliev Z S,  Ga$\beta$ S, Wolter A U B, Koroleva A V, Shikin A M, Blanco-Rey M, and \textit{et al}.
2019 Nature {\bf576} 416

\bibitem{W.Wang}
Wang W, Ou Y, Liu C, Wang Y, He K, Xue Q K, and Wu W, 
2018 Nat. Phys. {\bf10} 1038

\bibitem{J.A.Sobota}
Sobota J A, He Y, and Shen Z X, 
2021 Rev. Mod. Phys. {\bf93} 025006

\bibitem{J.Wu}
Wu J, Liu F, Sasase M, Ienaga K, Obata Y, Yukawa R, Horiba K, Kumigashira H,  Okuma S, Inoshita T, and Hosono H, 
2019 Sci. Adv. {\bf5} eaax9989

\bibitem{R.C.Vidal}
Vidal R C, Zeugner A, Facio J I, Ray R, Haghighi M H, Wolter A U B, Bohorquez L T C,  Caglieris F, Moser S, Figgemeier T, Peixoto T R F, and \textit{et al}.
2019 Phys. Rev. X {\bf9} 041065

\bibitem{I.I.Klimovskikh}
Klimovskikh I I, Otrokov M M, Estyunin D, Eremeev S V, Filnov S O, Koroleva A,
Shevchenko E, Voroshnin V, Rybkin A G, Rusinov I P, and \textit{et al},
2020 npj Quantum Mater. {\bf5} 54

\bibitem{M.Mogi}
Mogi M, Kawamura M, Tsukazaki A, Yoshimi R, Takahashi K S, Kawasaki M, Tokura Y, 
2017 Sci. Adv. {\bf3} eaao1669

\bibitem{D.Xiao}
Xiao D, Jiang J, Shin J H, Wang W, Wang F, Zhao Y F, Liu C, Wu W, Chan M H W, Samarth N, and Chang C Z, 
2018 Phys. Rev. Lett. {\bf120} 056801

\end{references}
\end{document}